\def\arxiv{}

\documentclass[runningheads]{llncs}

\usepackage{cite}
\usepackage{url}
\usepackage{amsmath}
\usepackage{algorithm}
\usepackage{algpseudocode}
\usepackage{enumitem}
\algrenewcommand\algorithmicrequire{\textbf{Input:}}
\algrenewcommand\algorithmicensure{\textbf{Output:}}

\makeatletter
\AtBeginDocument{%
  \def\doi#1{\url{https://doi.org/#1}}}
\makeatother

\usepackage{graphicx}
\usepackage[%
breaklinks=true,
colorlinks=true,
citecolor=blue,
linkcolor=blue,
urlcolor=blue,
]{hyperref}

\usepackage{siunitx}
\usepackage{acro}
\DeclareAcronym{hsi}{short=HSI,long=hyperspectral imaging}
\DeclareAcronym{dsc}{short=DSC,long=Dice similarity coefficient}
\DeclareAcronym{sd}{short=SD,long=standard deviation}
\DeclareAcronym{soa}{short=SOA,long=state-of-the-art}
\DeclareAcronym{ood}{short=OOD,long=out-of-distribution}
\DeclareAcronym{nsd}{short=NSD,long=normalized surface distance}

\usepackage{xspace}
\newcommand{\varTotalImages}{600\xspace}
\newcommand{\varTotalPigs}{33\xspace}

\newcommand{\varTotalPigsOld}{20\xspace}
\newcommand{\varTotalImagesIsolationReal}{94\xspace}
\newcommand{\varTotalPigsIsolationReal}{25\xspace}
\newcommand{\varTotalImagesGlove}{142\xspace}
\newcommand{\varTotalImagesNoGlove}{364\xspace}

\newcommand{\varTotalTrainingPigs}{15\xspace}
\newcommand{\varTotalTrainingImages}{340\xspace}
\newcommand{\varTotalTestPigsInDistribution}{5\xspace}
\newcommand{\varTotalTestImagesInDistribution}{166\xspace}
\newcommand{\varTotalClasses}{19\xspace}
\newcommand{\varTotalOrganClasses}{18\xspace}

\newcommand{\varDSCBaselineHSIInDistribution}{0.86 (\ac{sd} 0.10)\xspace}

\newcommand{\varDSCBaselineRGBInDistribution}{0.83 (\ac{sd} 0.10)\xspace}

\newcommand{\varRGBImprovementIsolationZeroClothMean}{\SI{61}{\percent}\xspace}
\newcommand{\varRGBImprovementIsolationReal}{\SI{46}{\percent}\xspace}
\newcommand{\varHSIImprovementIsolationZeroClothMean}{\SI{57}{\percent}\xspace}
\newcommand{\varHSIImprovementIsolationReal}{\SI{50}{\percent}\xspace}
\newcommand{\varHSIDropRange}{\SIrange{5}{45}{\percent}\xspace}
\newcommand{\varRGBDropRange}{\SIrange{10}{46}{\percent}\xspace}
\newcommand{\varHSIImprovementRangeDSC}{\SIrange{9}{90}{\percent}\xspace}
\newcommand{\varRGBImprovementRangeDSC}{\SIrange{9}{67}{\percent}\xspace}
\newcommand{\varHSIImprovementRangeNSD}{\SIrange{16}{96}{\percent}\xspace}
\newcommand{\varRGBImprovementRangeNSD}{\SIrange{15}{79}{\percent}\xspace}
\newcommand{\varHSIImprovementMax}{\SI{90}{\percent}\xspace}
\newcommand{\varRGBImprovementMax}{\SI{67}{\percent}\xspace}
\newcommand{\varHSIMaxDrop}{\SI{45}{\percent}\xspace}
\newcommand{\varRGBMaxDrop}{\SI{46}{\percent}\xspace}
\newcommand{\varRemovalLiverGallbladderDrop}{\SI{63}{\percent}\xspace}

\newcommand{\varHSIImprovementOcclusionPancreas}{\SI{283}{\percent}\xspace}
\newcommand{\varHSIImprovementOcclusionStomach}{\SI{69}{\percent}\xspace}

\newcommand{\dataset}[1]{\textit{#1}}

\ifx\arxiv\undefined
\newcommand{\citesuppl}[2]{#2}
\else
\newcommand{\citesuppl}[2]{\autoref{#1}}
\fi

\begin{document}
\title{Semantic segmentation of surgical hyperspectral images under geometric domain shifts}
\titlerunning{Semantic segmentation under geometric domain shifts}

\author{
Jan Sellner\inst{1,2,3,4,}\thanks{Equal contribution.} \and
Silvia Seidlitz\inst{1,2,3,4,\star} \and
Alexander Studier-Fischer\inst{4,5,6} \and
Alessandro Motta\inst{1} \and
Berkin Özdemir\inst{5,6} \and
Beat Peter Müller-Stich\inst{5,6} \and
Felix Nickel\inst{2,5,6} \and
Lena Maier-Hein\inst{1,2,3,4,6}
}
%
\authorrunning{Sellner \& Seidlitz et al.}
%
\institute{
Division of Intelligent Medical Systems (IMSY), German Cancer Research Center (DKFZ), Heidelberg, Germany \and
Helmholtz Information and Data Science School for Health, Karlsruhe/Heidelberg, Germany \and
Faculty of Mathematics and Computer Science, Heidelberg University, Heidelberg, Germany \and
National Center for Tumor Diseases (NCT), NCT Heidelberg, a partnership between DKFZ and university medical center Heidelberg \and
Department of General, Visceral, and Transplantation Surgery, Heidelberg University Hospital, Heidelberg, Germany \and
Medical Faculty, Heidelberg University, Heidelberg, Germany
}

\maketitle              
\begin{abstract}
Robust semantic segmentation of intraoperative image data could pave the way for automatic surgical scene understanding and autonomous robotic surgery. Geometric domain shifts, however -- although common in real-world open surgeries due to variations in surgical procedures or situs occlusions -- remain a topic largely unaddressed in the field. To address this gap in the literature, we (1) present the first analysis of \ac{soa} semantic segmentation networks in the presence of geometric \ac{ood} data, and (2) address generalizability with a dedicated augmentation technique termed 'Organ Transplantation' that we adapted from the general computer vision community. According to a comprehensive validation on six different \ac{ood} data sets comprising \varTotalImages RGB and \ac{hsi} cubes from \varTotalPigs pigs semantically annotated with \varTotalClasses classes, we demonstrate a large performance drop of \ac{soa} organ segmentation networks applied to geometric \ac{ood} data. Surprisingly, this holds true not only for conventional RGB data (drop of \ac{dsc} by \varRGBMaxDrop) but also for \ac{hsi} data (drop by \varHSIMaxDrop), despite the latter’s rich information content per pixel. Using our augmentation scheme improves on the \ac{soa} \ac{dsc} by up to \varRGBImprovementMax (RGB) and \varHSIImprovementMax (\ac{hsi}) and renders performance on par with in-distribution performance on real \ac{ood} test data. The simplicity and effectiveness of our augmentation scheme makes it a valuable network-independent tool for addressing geometric domain shifts in semantic scene segmentation of intraoperative data. Our code and pre-trained models are available at \href{https://github.com/IMSY-DKFZ/htc}{https://github.com/IMSY-DKFZ/htc}.
\end{abstract}
\acresetall
\section{Introduction}
\begin{figure}
    \centering
    \includegraphics[width=\linewidth]{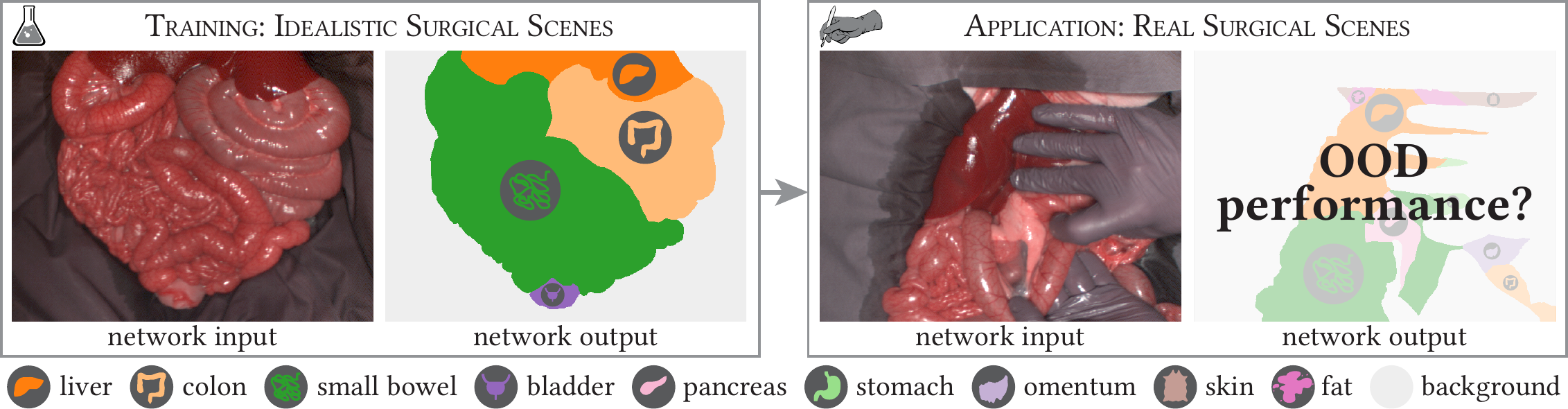}
    \caption{\Acf*{soa} surgical scene segmentation networks show promising results on idealistic datasets. However, in real-world surgeries, geometric domain shifts such as occlusions of the situs by operating staff are common. The generalizability of \acs*{soa} algorithms towards geometric \acf*{ood} has not yet been addressed.}
    \label{fig:problem}
\end{figure}

Automated surgical scene segmentation is an important prerequisite for context-aware assistance and autonomous robotic surgery. Recent work showed that deep learning-based surgical scene segmentation can be achieved with high accuracy~\cite{kadkhodamohammadi_feature_2019, scheikl_deep_2020} and even reach human performance levels if using \ac{hsi} instead of RGB data, with the additional benefit of providing functional tissue information \cite{seidlitz_robust_2022}. However, to our knowledge, the important topic of geometric domain shifts commonly present in real-world surgical scenes (e.g., situs occlusions, cf. \autoref{fig:problem}) so far remains unaddressed in literature. It is questionable whether the \ac{soa} image-based segmentation networks in \cite{seidlitz_robust_2022} are able to generalize towards an \ac{ood} context. The only related work by Kitaguchi et al.~\cite{kitaguchi_limited_2022} showed that surgical instrument segmentation algorithms fail to generalize towards unseen surgery types that involve known instruments in an unknown context. We are not aware of any investigation or methodological contribution on geometric domain shifts in the context of surgical scene segmentation.

Generalizability in the presence of domain shifts is being intensively studied by the general machine learning community. Here, data augmentation evolved as a simple, yet powerful technique \cite{shorten_survey_2019, alomar_data_2023}. 
In deep learning-based semantic image segmentation, geometric transformations are most common \cite{kar_review_2021}. This holds particularly true for surgical applications. Our analysis of the \ac{soa} (35 publications on tissue or instrument segmentation) exclusively found geometric (e.g., rotating), photometric (e.g., color jittering) and kernel (e.g., Gaussian blur) transformations and only in a single case elastic transformations and Random Erasing (within an image, a rectangular area is blacked out) \cite{zhong_random_2020} being applied. Similarly, augmentations in \ac{hsi}-based tissue classification are so far limited to geometric transformations. To our knowledge, the potential benefit of complementary transformations proposed for image classification and object detection, such as Hide-and-Seek (an image is divided into a grid of patches that are randomly blacked out) \cite{singh_hide-and-seek_2017}, Jigsaw (images are divided into a grid of patches and patches are randomly exchanged between images) \cite{chen_image_2019}, CutMix (a rectangular area is copied from one image onto another image) \cite{yun_cutmix_2019} and CutPas (an object is placed onto a random background scene) \cite{dwibedi_cut_2017} (cf. Fig. 2), remains unexplored.

Given these gaps in the literature, the contribution of this paper is twofold:
\begin{enumerate}
    \item We show that geometric domain shifts have disastrous effects on \ac{soa} surgical scene segmentation networks for both conventional RGB and \ac{hsi} data.
    \item We demonstrate that topology-altering augmentation techniques adapted from the general computer vision community are capable of addressing these domain shifts.
\end{enumerate}

\begin{figure}
    \centering
    \includegraphics[width=\linewidth]{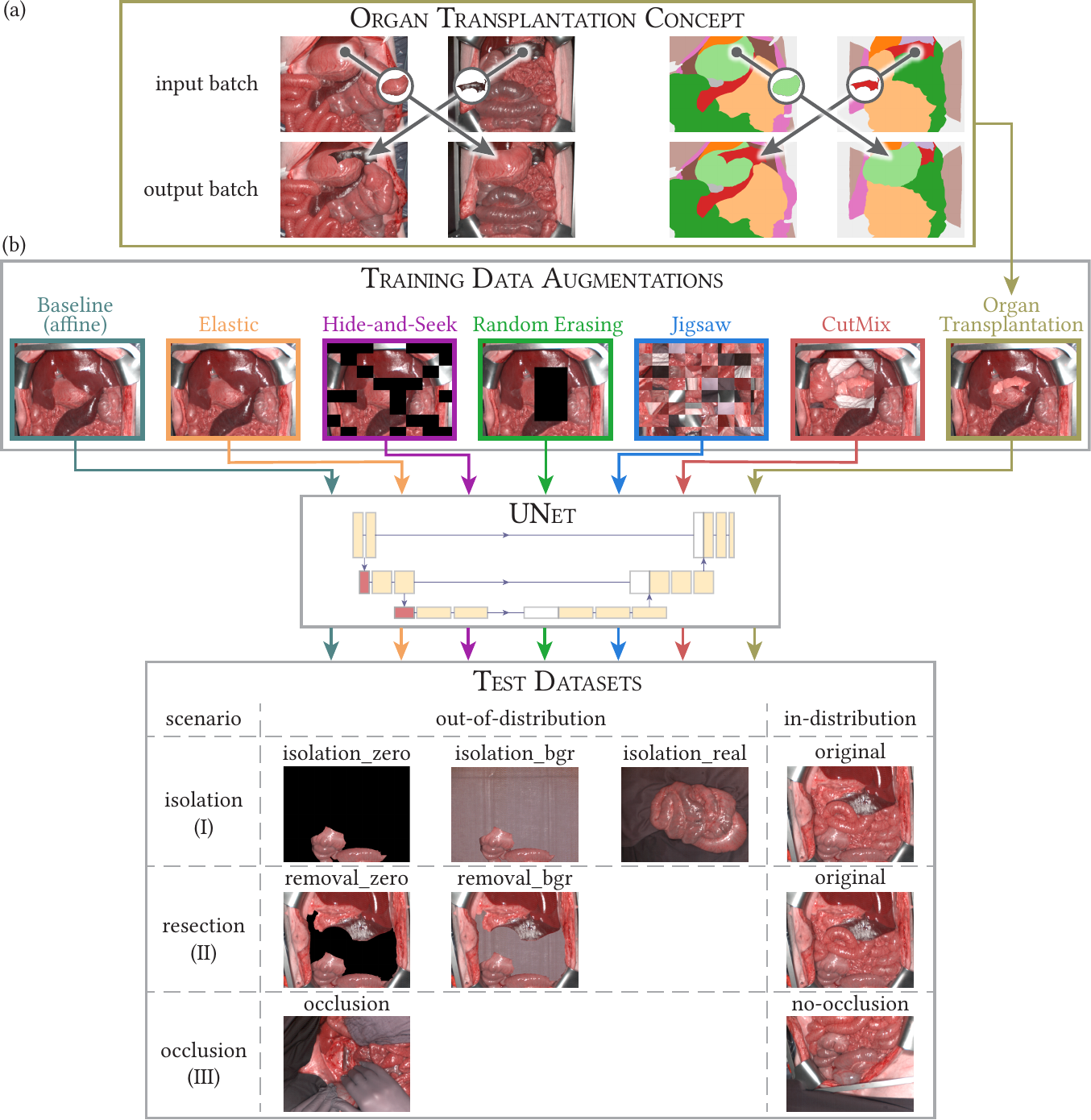}
    \caption{(a) \textit{Organ Transplantation} augmentation concept inspired from~\cite{dwibedi_cut_2017}. Image features and corresponding segmentations of randomly selected organs are transferred between images in one batch (in the example, the stomach is transferred from the left to the right and the spleen from the right to the left image). (b) Illustration of our validation experiments. We assess the generalizability under geometric domain shifts of seven different data augmentation techniques in deep learning-based organ segmentation. We validate the model performance on a range of \acf*{ood} scenarios, namely (1) organs in isolation (\dataset{isolation\_zero}, \dataset{isolation\_bgr} and \dataset{isolation\_real}), (2) organ resections (\dataset{removal\_zero} and \dataset{removal\_bgr}), and (3) situs occlusions (\dataset{occlusion}), in addition to in-distribution data (\dataset{original} and \dataset{no-occlusion} (subset of \dataset{original} without occlusions)).}
    \label{fig:overview}
\end{figure}

\section{Materials and methods}
The following sections describe the network architecture, training setup and augmentation methods (\autoref{sec:networks}), and our experimental design, including an overview of our acquired datasets and validation pipeline (\autoref{sec:experiments}).

\subsection{Deep learning-based surgical scene segmentation}
\label{sec:networks}

Our contribution is based on the assumption that application-specific data augmentation can potentially address geometric domain shifts. Rather than changing the network architecture of previously successful segmentation methods, we adapt the data augmentation. 

\textit{Surgery-inspired augmentation:} Our Organ Transplantation augmentation illustrated in \autoref{fig:overview} has been inspired by the image-mixing augmentation CutPas that was originally proposed for object detection \cite{dwibedi_cut_2017} and recently adapted for instance segmentation \cite{ghiasi_simple_2021} and low-cost dataset generation via image synthesis from few real-world images in surgical instrument segmentation \cite{wang_rethinking_2022}. It is based on placing an organ into an unusual context while keeping shape and texture consistent. This is achieved by transplanting all pixels belonging to one object class (e.g., an organ class or background) into a different surgical scene.
Our selection of further computer vision augmentation methods that could potentially improve geometric \ac{ood} performance (cf. \autoref{fig:overview}) was motivated by the specific conditions encountered in surgical procedures (cf. \autoref{sec:experiments} for an overview). The noise augmentations Hide-and-Seek and Random Erasing black out all pixels inside rectangular regions within an image, thereby generating artificial situs occlusions. Instead of blacking out, the image-mixing techniques Jigsaw and CutMix copy all pixels inside rectangular regions within an image into a different surgical scene. We adapted the image-mixing augmentations to our segmentation task by also copying and pasting the corresponding segmentations. Hence, apart from occluding the underlying situs, image parts/organs occur in an unusual neighborhood.

\textit{Network architecture and training:} We used a U-Net architecture \cite{ronneberger_u-net_2015} with an efficientnet-b5 encoder \cite{tan_efficientnet_2019} pre-trained on ImageNet data and using stochastic weight averaging \cite{izmailov_averaging_2018} for both RGB and \ac{hsi} data as it achieved human performance level in recent work~\cite{seidlitz_robust_2022}. As a pre-processing step, the \ac{hsi} data was calibrated with white and dark reference images and $\ell^1$-normalized to remove the influence of multiplicative illumination changes. Dice and cross-entropy loss were equally weighted to compute the loss function. The Adam optimization algorithm \cite{kingma_adam_2017} was used with an exponential learning rate scheduler. Training was performed for 100 epochs with a batch size of five images.

\subsection{Experiments}
\label{sec:experiments}

To study the performance of \ac{soa} surgical scene segmentation networks under geometric domain shifts and investigate the generalizability improvements offered by augmentation techniques, we covered the following \ac{ood} scenarios:

\begin{enumerate}[label=(\Roman*),align=left,leftmargin=6ex,labelwidth=5ex,labelsep=0ex]
    \item \dataset{Organs in isolation}: Abdominal linens are commonly used  to protect soft tissue and organs, counteract excessive bleeding, and absorb blood and secretion. Some surgeries (e.g., enteroenterostomy), even require covering all but a single organ. In such cases, an organ needs to be robustly identified without any information on neighboring organs.
    \item \dataset{Organ resections}: In resection procedures, parts or even the entirety of an organ are removed and surrounding organs thus need to be identified despite the absence of a common neighbor.
    \item \dataset{Occlusions}: Large parts of the situs can be occluded by the surgical procedure itself, introducing \ac{ood} neighbors (e.g., gloved hands). The non-occluded parts of the situs need to be correctly identified.
\end{enumerate}

\textit{Real-world datasets:} In total, we acquired \varTotalImages intraoperative \ac{hsi} cubes from \varTotalPigs pigs using the \ac{hsi} system Tivita\textsuperscript{\textregistered} Tissue (Diaspective Vision GmbH, Am Salzhaff, Germany). These were semantically annotated  with  background and \varTotalOrganClasses tissue classes, namely heart, lung, stomach, small intestine, colon, liver, gallbladder, pancreas, kidney with and without Gerota's fascia, spleen, bladder, subcutaneous fat, skin, muscle, omentum, peritoneum, and major veins. Each \ac{hsi} cube captures 100 spectral channels in the range between \SI{500}{nm} and \SI{1000}{nm} at an image resolution of $640\times480$ pixels. RGB images were reconstructed by aggregating spectral channels in the blue, green, and red ranges. To study organs in isolation, we acquired \varTotalImagesIsolationReal images from \varTotalPigsIsolationReal pigs in which all but a specific organ were covered by abdominal linen for all \varTotalOrganClasses different organ classes (dataset \dataset{isolation\_real}). To study the effect of occlusions, we acquired \varTotalImagesGlove images of \varTotalPigsOld pigs with real-world situs occlusions (dataset \dataset{occlusion}), and \varTotalImagesNoGlove occlusion-free images (dataset \dataset{no-occlusion}). Example images are shown in \autoref{fig:overview}.

\textit{Manipulated data:} We complemented our real-world datasets with four manipulated datasets. To simulate organs in isolation, we replaced every pixel in an image $I$ that does not belong to the target label $l$ either with zeros or spectra copied from a background image. We applied this transformation to all images in the dataset \dataset{original} and all target labels $l$, yielding the datasets \dataset{isolation\_zero} and \dataset{isolation\_bgr}. Similarly, we simulated organ resections by replacing all pixels belonging to the target label $l$ either with zeros or background spectra, yielding the datasets \dataset{removal\_zero} and \dataset{removal\_bgr}. Example images are shown in \autoref{fig:overview}.

\textit{Train-test split and hyperparameter tuning:} The \ac{soa} surgical scene segmentation algorithms are based on a union of the datasets \dataset{occlusion} and \dataset{no-occlusion}, termed dataset \dataset{original}, which was split into a hold-out test set (\varTotalTestImagesInDistribution images from \varTotalTestPigsInDistribution pigs) and a training set (\varTotalTrainingImages images from \varTotalTrainingPigs pigs). To enable a fair comparison, the same train-test split on pig level was used across all networks and scenarios. This also holds for the occlusion scenario, in which the dataset \dataset{no-occlusion} was used instead of \dataset{original} for training. All networks used the geometric transformations shift, scale, rotate, and flip from the \ac{soa} prior to applying the augmentation under examination. All hyperparameters were set according to the \ac{soa}. Only hyperparameters related to the augmentation under examination, namely the probability $p$ of applying the augmentation, were optimized through a grid search with $p \in \{ 0.2, 0.4, 0.6, 0.8, 1 \}$. We used five-fold-cross-validation on the datasets \dataset{original}, \dataset{isolation\_zero}, and \dataset{isolation\_bgr} to tune $p$ such that good segmentation performance was achieved on both in-distribution and \ac{ood} data.

\textit{Validation strategy:} Following the recommendations of the Metrics Reloaded framework \cite{maier-hein_metrics_2023}, we combined the \ac{dsc} \cite{dice_measures_1945} as an overlap-based metric with the boundary-based metric \ac{nsd} \cite{nikolov_deep_2021} for validation for each class $l$. To respect the hierarchical test set structure, metric aggregation was performed by first macro-averaging the class-level metric value $M_{l}$ ($M \in \{ \operatorname{DSC}, \operatorname{NSD}\}$) across all images of one pig and subsequently across pigs. The organ removal experiment required special attention in this context, as multiple $M_{l}$ values per image could be generated corresponding to all the possible neighbour organs that could be removed. In this case, we selected for each $l$ the minimum of all $M_{l}$ values, which corresponds to the segmentation performance obtained after removing the most important neighbour of $l$. The same class-specific \ac{nsd} thresholds as in the \ac{soa} were used.

\begin{figure}[ht]
    \centering
    \includegraphics[width=\linewidth]{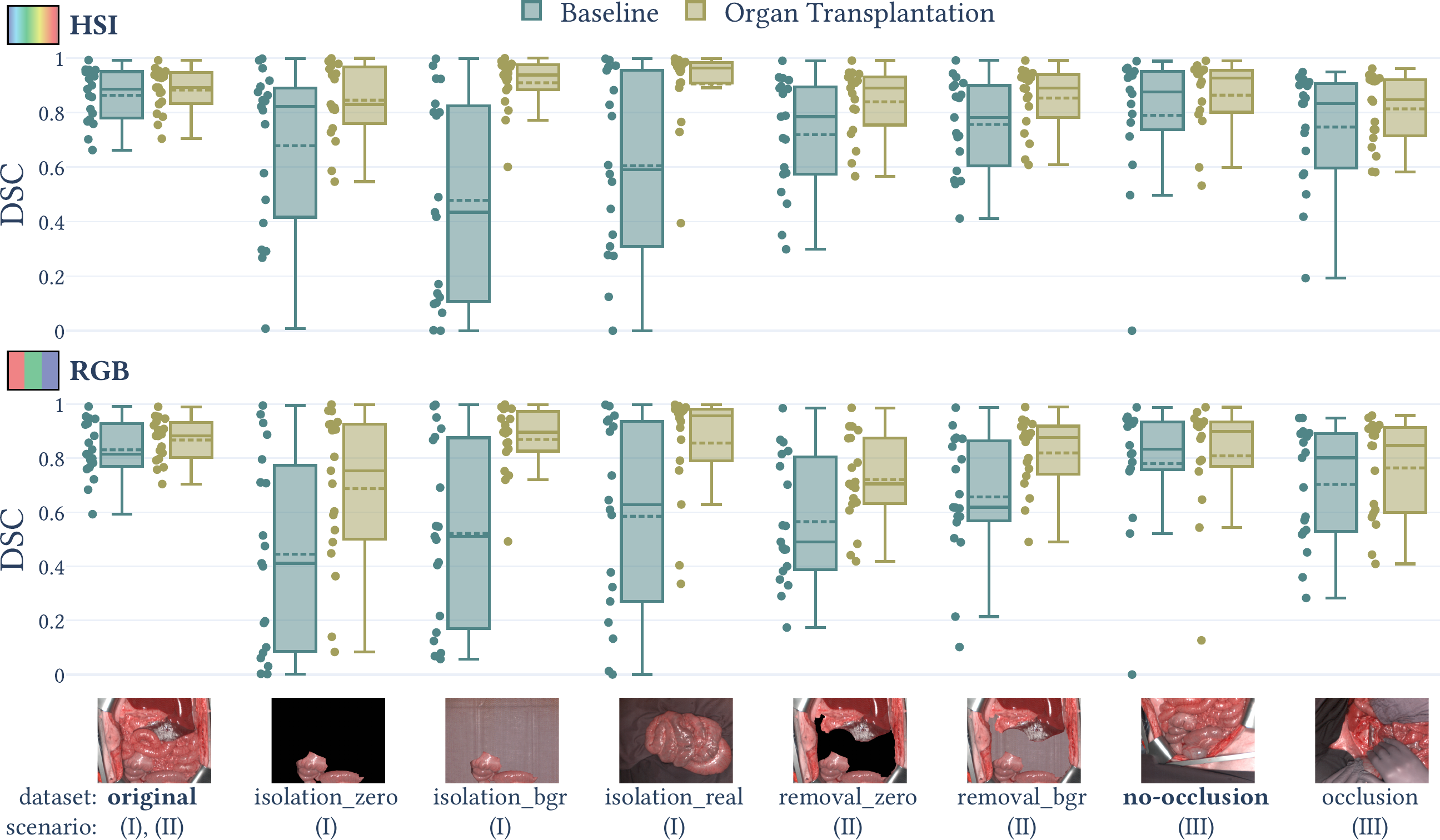}
    \caption{Segmentation performance of the \acf*{hsi} and RGB modality for all eight test datasets (six \acf*{ood} and two in-distribution datasets (bold)) comparing the baseline network with the Organ Transplantation network. Each point denotes one out of 19 class-level \acf*{dsc} values after hierarchical aggregation across images and subjects. The boxplots show the quartiles of the class-level \ac{dsc}. The whiskers extend up to 1.5 times the interquartile range and the median and mean are represented as a solid and dashed line, respectively.}
    \label{fig:boxplots}
\end{figure}

\section{Results}

\textit{Effects of geometric domain shifts:} When applying a \ac{soa} segmentation network to geometric \ac{ood} data, the performance drops radically (cf. \autoref{fig:boxplots}). Starting from a high \ac{dsc} for in-distribution data (RBG: \varDSCBaselineRGBInDistribution; HSI: \varDSCBaselineHSIInDistribution), the performance drops by \varRGBDropRange for RGB and by \varHSIDropRange for \ac{hsi}, depending on the experiment. In the organ resection scenario, the largest drop in performance of \varRemovalLiverGallbladderDrop occurs for the gallbladder upon liver removal (cf. \citesuppl{fig:cm_removal}{Suppl. Fig. 1}). Similar trends can be observed for the boundary-based metric \ac{nsd}, as shown in \citesuppl{fig:boxplots_NSD}{Suppl. Fig. 2}.

\textit{Performance of our method:} \autoref{fig:boxplots} and \citesuppl{fig:boxplots_NSD}{Suppl. Fig. 2} show that the Organ Transplantation augmentation (gold) can address geometric domain shifts for both the RGB and \ac{hsi} modality. The latter yields consistently better results, indicating that the spectral information is crucial in situations with limited context. The performance improvement compared to the baseline ranges from \varRGBImprovementRangeDSC (\ac{dsc}) and \varRGBImprovementRangeNSD (\ac{nsd}) for RGB, and from \varHSIImprovementRangeDSC (\ac{dsc}) and \varHSIImprovementRangeNSD (\ac{nsd}) for \ac{hsi}, with the benefit on \ac{ood} data being largest for organs in isolation and smallest for situs occlusions. The Organ Transplantation augmentation even slightly improves performance on in-distribution data (\dataset{original} and \dataset{no-occlusion}). Upon encountering situs occlusions, the largest \ac{dsc} improvement is obtained for the organ classes pancreas (\varHSIImprovementOcclusionPancreas) and stomach (\varHSIImprovementOcclusionStomach). For organs in isolation, the performance improvement on manipulated data (\ac{dsc} increased by \varHSIImprovementIsolationZeroClothMean (\ac{hsi}) and \varRGBImprovementIsolationZeroClothMean (RGB) on average) is comparable to that on real data (\ac{dsc} increased by \varHSIImprovementIsolationReal (\ac{hsi}) and \varRGBImprovementIsolationReal (RGB)).

\begin{figure}[!ht]
    \centering
    \includegraphics[width=\linewidth]{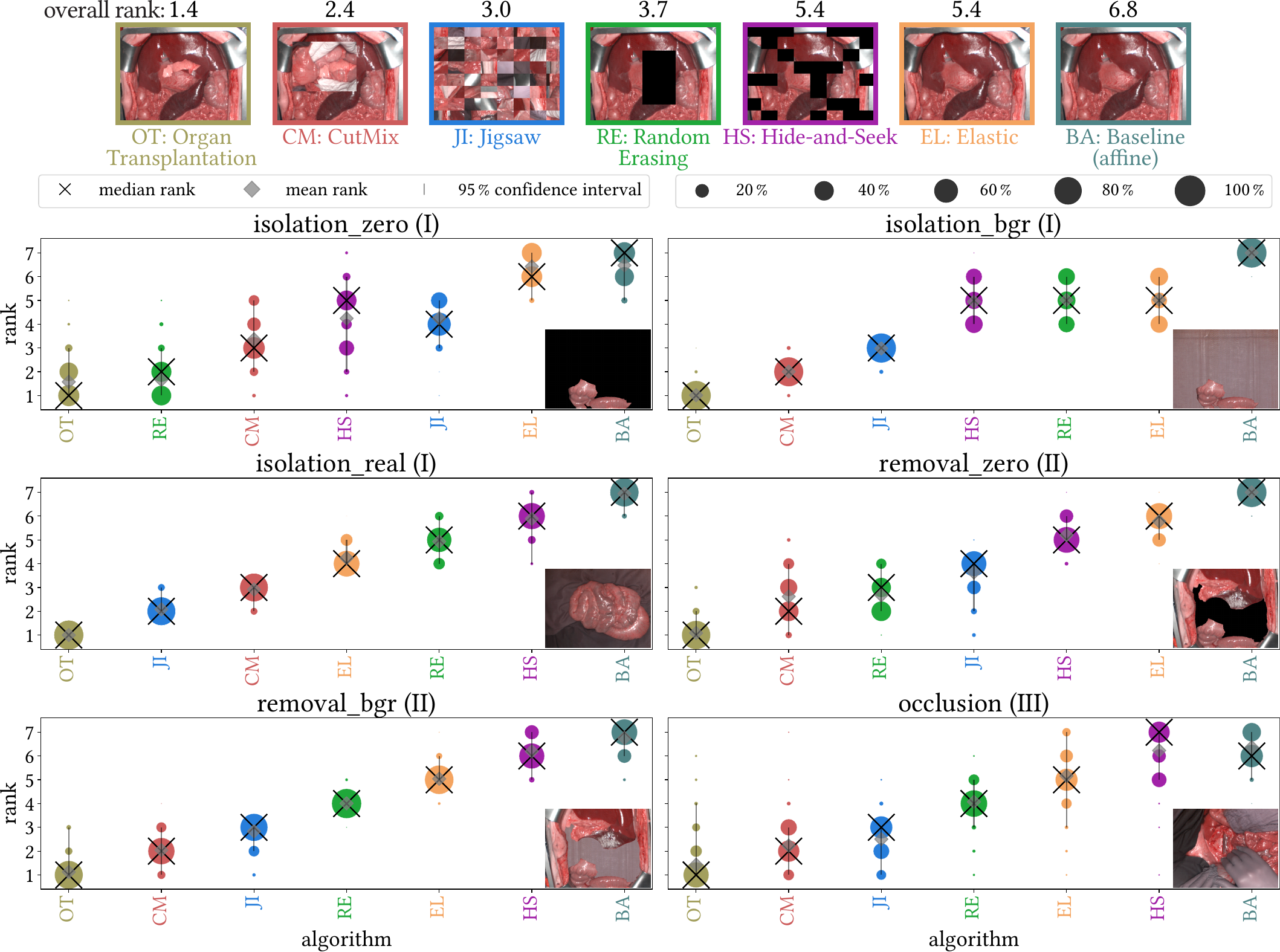}
    \caption{Uncertainty-aware ranking of the seven augmentation methods for all six geometric \acf*{ood} test datasets. Organ Transplantation consistently ranks first and baseline last. The area of each blob for one rank and algorithm is proportional to the relative frequency of that algorithm achieving the respective rank across 1000 bootstrap samples consisting of \varTotalClasses hierarchically aggregated class-level \acf*{dsc} values each (concept from \cite{wiesenfarth_methods_2021}). The numbers above the example images denote the overall ranking across datasets (mean of all mean ranks).}
    \label{fig:ranking_bubbleplot}
\end{figure}

\textit{Comparison to \ac{soa} augmentations:} There is no consistent ranking across all six \ac{ood} datasets except for Organ Transplantation always ranking first and baseline usually ranking last (cf. \autoref{fig:ranking_bubbleplot} for \ac{dsc}- and \citesuppl{fig:ranking_bubbleplot_NSD}{Suppl. Fig. 3} for \ac{nsd}-based ranking). Overall, image-mixing augmentations outperform noise augmentations. Augmentations that randomly sample rectangles usually rank better than comparable augmentations using a grid structure (e.g., CutMix vs. Jigsaw). 

\section{Discussion}
To our knowledge, we are the first to show that \ac{soa} surgical scene segmentation networks fail under geometric domain shifts. We were particularly surprised by the large performance drop for \ac{hsi} data, rich in spectral information. Our results clearly indicate that \ac{soa} segmentation models rely on context information. 

Aiming to address the lack of robustness to geometric variations, we adapted so far unexplored topology-altering data augmentation schemes to our target application and analyzed their generalizability on a range of six geometric \ac{ood} datasets specifically designed for this study. The Organ Transplantation augmentation outperformed all other augmentations and resulted in similar performance to in-distribution performance on real \ac{ood} data. Besides its effectiveness and computational efficiency, we see a key advantage in its potential to reduce the amount of real \ac{ood} data required in network training. Our augmentation networks were optimized on simulated \ac{ood} data, indicating that image manipulations are a powerful tool for judging geometric \ac{ood} performance if real data is unavailable, such as in our resection scenario, which would have required an unfeasible number of animals. With laparoscopic \ac{hsi} systems only recently becoming available, the investigation and compensation of geometric domain shifts in minimally-invasive surgery could become a key direction for future research. Our proposed augmentation is model-independent, computationally efficient and effective, and thus a valuable tool for addressing geometric domain shifts in semantic scene segmentation of intraoperative \ac{hsi} and RGB data. Our implementation and models will be made publicly available.


\section*{Acknowledgements and data usage}
This project was supported by the European Research Council (ERC) under the European Union’s Horizon 2020 research and innovation programme (NEURAL SPICING, 101002198), the National Center for Tumor Diseases (NCT) Heidelberg's Surgical Oncology Program, the German Cancer Research Center (DKFZ), and the Helmholtz Association under the joint research school HIDSS4Health (Helmholtz Information and Data Science School for Health). The private \ac{hsi} data was acquired at Heidelberg University Hospital after approval by the Committee on Animal Experimentation (G-161/18 and G-262/19).

%
%
%
\bibliographystyle{splncs04}
\bibliography{references_concise}

\ifx\arxiv\undefined
\else
\newpage
\appendix

\section{Supplementary material}

\begin{figure}[!htb]
    \centering
    \includegraphics[width=\linewidth]{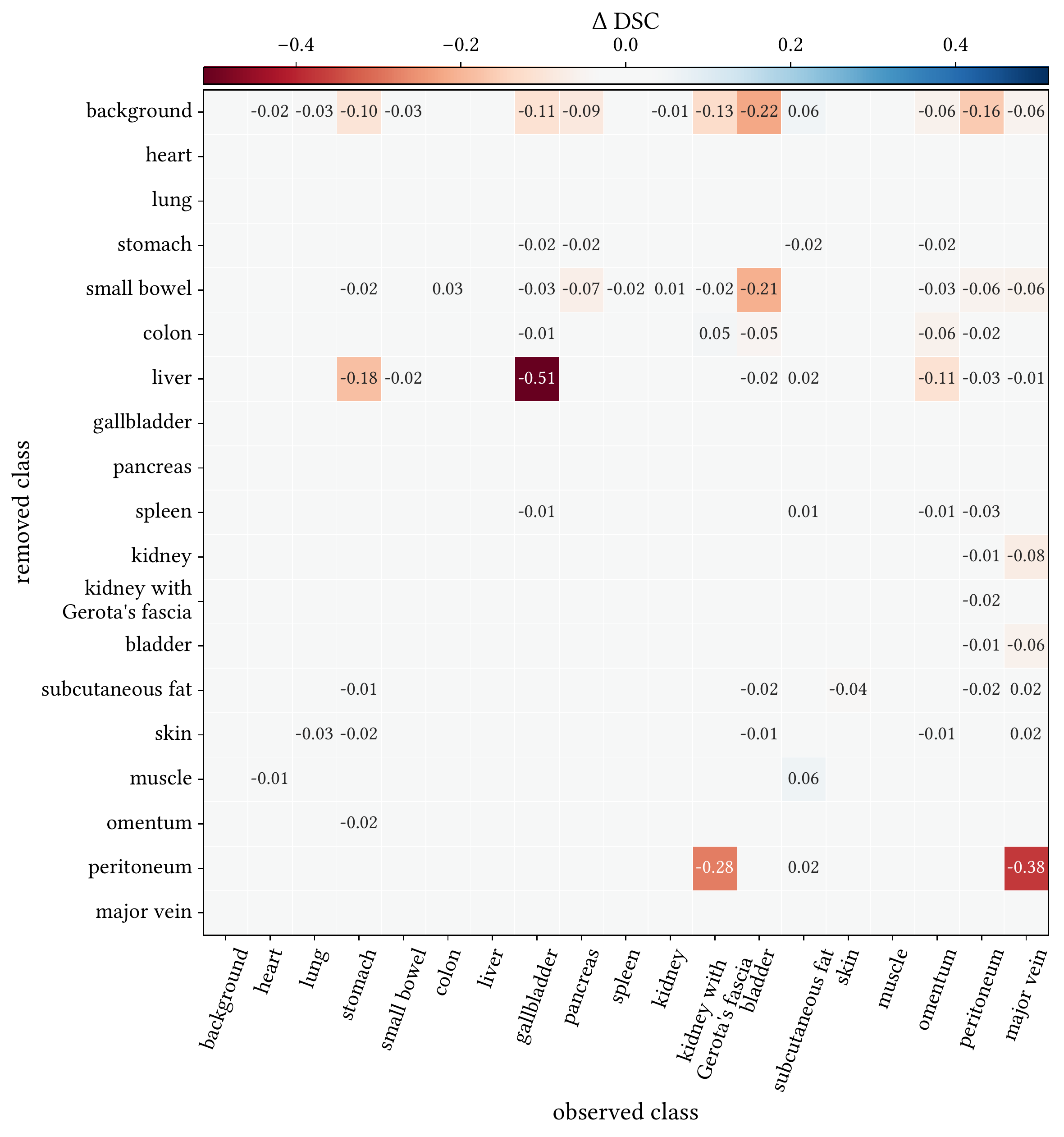}
    \caption{Drop in performance for the baseline network on hyperspectral images upon encountering organ removals, stratified by observed (columns) and removed (row) class. The $(i, j)$-th entry in the matrix depicts the change in \acf*{dsc} of the $j$-th organ when the $i$-th organ is removed in the images. The $\Delta \operatorname{DSC}$ values were computed for every organ in an image and hierarchically aggregated across images and subjects. Values of $|\Delta \operatorname{DSC}|<0.01$ are not shown for clarity.}
    \label{fig:cm_removal}
\end{figure}

\begin{figure}[htb]
    \centering
    \includegraphics[width=\linewidth]{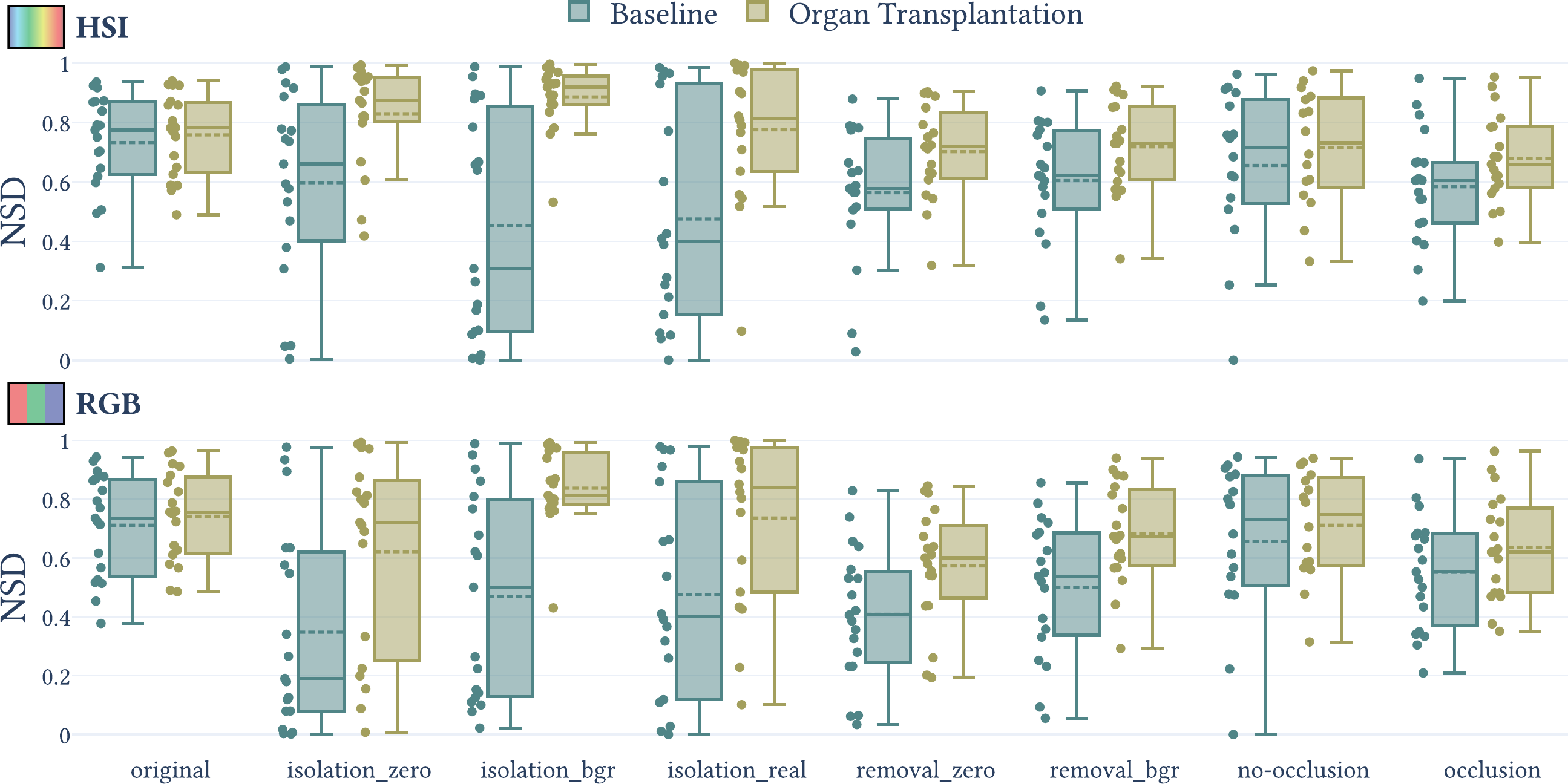}
    \caption{Segmentation performance of the baseline and Organ Transplantation networks on hyperspectral and RGB images. Equivalent to Fig. 3 using the \acf*{nsd} (thresholds from \cite{seidlitz_robust_2022}) instead of the \acf*{dsc}.}
    \label{fig:boxplots_NSD}
\end{figure}

\begin{figure}[htb]
    \centering
    \includegraphics[width=\linewidth]{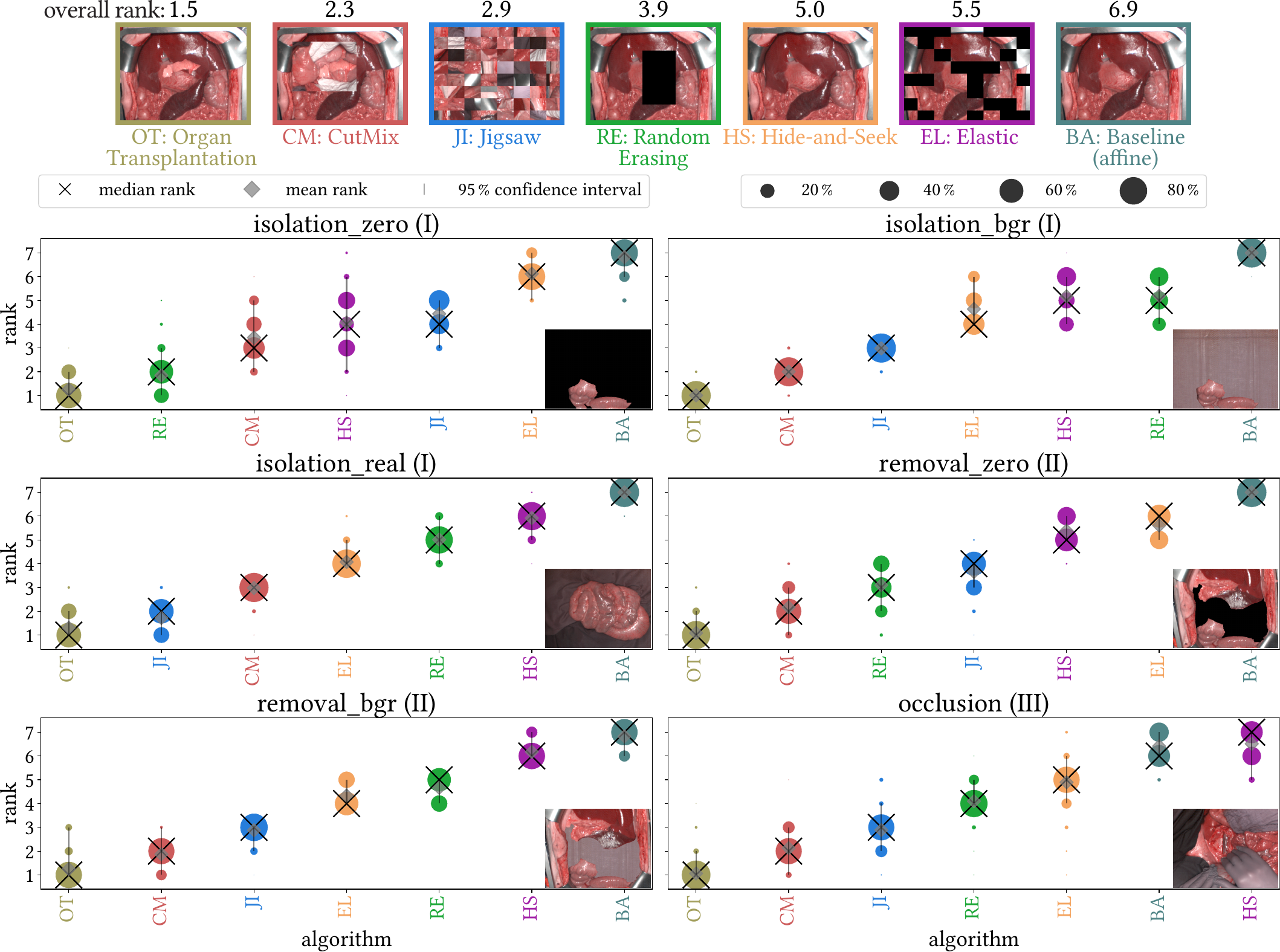}
    \caption{Uncertainty-aware ranking of the augmentation methods on geometric \acf*{ood} hyperspectral images. Equivalent to Fig. 4 using the \acf*{nsd} (thresholds from \cite{seidlitz_robust_2022}) instead of the \acf*{dsc}.}
    \label{fig:ranking_bubbleplot_NSD}
\end{figure}
\fi

\end{document}